\documentclass[twocolumn,showpacs,floats,floatfix,superscriptaddress,aps,pra]{revtex4}
\usepackage{amsfonts}
\usepackage{amssymb}
\usepackage{amsmath}
\usepackage{calc,graphicx,color,bm}

\def\be{ \begin{equation} }
\def\ee{ \end{equation} }
\def\bea{ \begin{eqnarray} }
\def\eea{ \end{eqnarray} }
\def\bse{ \begin{subequations} }
\def\ese{ \end{subequations} }
\def\sech{\,\text{sech}\,}

\def\i{\,i\,}
\def\e{\,e\,}
\def\const{\,\text{const}}

\def\t{(t)}
\def\dt{\frac{\d}{\d t}}
\def\dt{\partial_t}

\def\to{\rightarrow}

\def\half{\tfrac12}

\newcommand{\ket}[1]{\vert #1\rangle}

\def\H{\mathbf{H}}

\def\c{\mathbf{c}}

\def\i{{\emph{i}}}

\def\e{{\emph{e}}}

\def\csch{\,\text{csch}}

\def\ibid{\emph{ibid.} }
\def\etal{\emph{et al.}}

\begin{document}

\author{Boyan T. Torosov}
\altaffiliation{Permanent address: Institute of Solid State Physics, Bulgarian Academy of Sciences, 72 Tsarigradsko chauss\'{e}e, 1784 Sofia, Bulgaria}
\affiliation{Dipartimento di Fisica, Politecnico di Milano and Istituto di Fotonica e Nanotecnologie del Consiglio Nazionale delle Ricerche, Piazza L. da Vinci 32, I-20133 Milano, Italy}
\author{Giuseppe Della Valle}
\affiliation{Dipartimento di Fisica, Politecnico di Milano and Istituto di Fotonica e Nanotecnologie del Consiglio Nazionale delle Ricerche, Piazza L. da Vinci 32, I-20133 Milano, Italy}
\author{Stefano Longhi}
\affiliation{Dipartimento di Fisica, Politecnico di Milano and Istituto di Fotonica e Nanotecnologie del Consiglio Nazionale delle Ricerche, Piazza L. da Vinci 32, I-20133 Milano, Italy}
\title{Non-Hermitian shortcut to adiabaticity}
\date{\today }

\begin{abstract}
A non-Hermitian shortcut to adiabaticity is introduced. By adding an imaginary term in the diagonal elements of the Hamiltonian of a two state quantum system, we show how one can cancel the nonadiabatic losses and perform an arbitrarily fast population transfer, without the need to increase the coupling. We apply this technique to two popular level-crossing models: the Landau-Zener model and the Allen-Eberly model.
\end{abstract}


\maketitle

%%%%%%%%%%%%%%%%%%%%%%%%%%%%%%%%%%%%%%%%%%%%%%%%%%%%%%%%%%%%%%%%%%%%%%%%%%%%%%%%%%%%%%%%%%%%%%%%%%%%%%%%%%%%%%%%%%%%%%%
\section{Introduction}
Techniques, based on adiabatic evolution are among the most popular methods for coherent atomic manipulation \cite{NVV-adiabatic}. During the adiabatic evolution, the system remains in an eigenstate of the time-dependent Hamiltonian. If this eigenstate is such that it connects different bare states in the beginning and in the end of the evolution, than it can be used for population transfer. However, perfectly adiabatic evolution is hard to realize, and in a real experiment usually there exists some (hopefully small) nonadiabatic coupling, which causes transitions between the eigenstates of the Hamiltonian and decreases the efficiency of the population transfer.
Many methods have been proposed to improve adiabatic evolution, for instance by using composite pulses \cite{CAP} or parallel adiabatic passage \cite{parallel}. Another popular approach is to use a \emph{shortcut to adiabaticity}, by adding extra fields which aim to nullify the nonadiabatic coupling \cite{shortcuts}. Most of the proposed techniques only use Hermitian Hamiltonians, and only little is known in the non-Hermitian case \cite{shortcuts-NH}. In the recent years, however, an increasing interest has been devoted to study non-Hermitian Hamiltonians, especially in the context of $\mathcal{PT}$-symmetric systems \cite{PT}. It was demonstrated, for instance, that a $\mathcal{PT}$-symmetric Hamiltonian can produce a faster than Hermitian evolution in a two-state quantum system, while keeping the eigenenergy difference fixed \cite{Faster}. Some non-Hermitian extensions \cite{LZ-NH} were also done to the Landau-Zener (LZ) model, which is a standard tool for the description of level-crossing systems. The dynamics of certain time-dependent $\mathcal{PT}$-symmetric two-level Hamiltonians have been discussed in recent works as well \cite{uff1,uff2}.
In this paper, we propose to realize a shortcut to adiabaticity by using complex-valued energies, which produce a non-hermitian Hamiltonian. Our approach allows to improve the speed of the rapid adiabatic passage, without the need to increase the coupling, unlike in the standard shortcuts methods in Hermitian systems. We apply this approach to two famous models, namely the LZ and the Allen-Eberly (AE) models, both of which are used for the description of level-crossing systems. The method is, however, applicable to a much larger variety of models. The main differences between Hermitian and non-Hermitian adiabatic shortcut methods will be discussed. 

\section{Adiabatic passage in a two-level system}

In this section we will briefly review the theory behind the rapid adiabatic passage for a two-level Hermitian system (see, e.g., \cite{NVV-adiabatic}).
The dynamics of a two-state quantum system is described by the Schr\"{o}dinger equation,
\be\label{Schr}
\i \hbar\partial_t \mathbf{c}(t) = \H(t)\mathbf{c}(t),
\ee
where the vector $\c(t) = [c_1(t), c_2(t)]^T$ contains the two probability amplitudes of the bare (diabatic) states $\ket{1}$ and $\ket{2}$.
The Hamiltonian in the rotating-wave approximation is
\be\label{H}
\H(t) = \frac\hbar 2 \left[ \begin{array}{cc} 0  & \Omega (t) \\
\Omega(t) & 2\Delta\t
\end{array} \right],
\ee
where $\Omega(t)$ is the Rabi frequency, which quantifies the strength of the coupling between states $\ket{1}$ and $\ket{2}$, and $\Delta(t)$ is the detuning between the external field and the Bohr frequency of the transition. 
To study adiabatic passage, we introduce the so-called adiabatic states $\ket{\varphi_+(t)}$ and $\ket{\varphi_-(t)}$, which are the eigenstates of the time-dependent Hamiltonian,
\be
\H(t) \ket{\varphi_\pm(t)} =\lambda_\pm \t \ket{\varphi_\pm(t)} ,  
\ee
with eigenvalues
\be\label{eigenenergies}
\lambda_\pm \t = \tfrac12 [\Delta\t \pm \sqrt{\Omega\t^2 + \Delta\t^2}].
\ee
The amplitudes in the adiabatic basis $\mathbf{a}\t=[a_{-}\t, a_{+}\t]^T$ are connected with the diabatic ones $\mathbf{c}\t$ via the rotation matrix
\be
\mathbf{R}(\theta ) = \left[ \begin{array}{cc}
\cos \theta  & \sin \theta  \\
-\sin \theta  & \cos \theta
\end{array}\right] ,
\ee
as $\mathbf{c}\t=\mathbf{R}(\theta\t)\mathbf{a}\t$, where $\theta\t =\frac{1}{2}\text{arctan}(\Omega /\Delta )$.
The Schr\"{o}dinger equation in the adiabatic basis reads
\be\label{Schr-adiabatic}
\i\hbar\dt \mathbf{a}\t=\mathbf{H}_{a}\t\mathbf{a}\t,
\ee
where
\be\label{H adiabatic}
\mathbf{H}_a \t = \hbar\left[ \begin{array}{cc}
\lambda_-\t & -\i\dot{\theta}\t \\
\i\dot{\theta}\t & \lambda_+\t
\end{array} \right] .
\ee
If $|\dot{\theta}\t| \ll \lambda_+\t - \lambda_-\t=\lambda\t$, then the evoution is adiabatic and we can neglect the transitions between the adiabatic states. Finally, if $\Omega\t$ and $\Delta\t$ are chosen in such way that 
\be
\lim_{t\to -\infty} \theta\t \rightarrow \pi/2 , \quad \lim_{t\to \infty} \theta\t \rightarrow 0
\ee
then we have
\be
\ket{\varphi_+(-\infty)}= \ket{1},\quad \ket{\varphi_+(\infty)}= \ket{2},
\ee
which means that the adiabatic following will produce complete population transfer from state $\ket{1}$ to $\ket{2}$. The efficiency of this transfer is limited by the adiabatic condition $|\dot{\theta}\t| \ll \lambda\t$, which requires slow evolution. Complete population transfer by adiabatic following can be realized, for instance by using the LZ model \cite{LZ}, the AE model \cite{AE}, or by using linearly chirped Gaussian pulses \cite{GSV}.

\section{Non-Hermitian shortcuts}
In this section we will show how the efficiency of the transfer can be improved by adding a suitably chosen non-Hermitian term $\i\gamma\t$, which aims to nullify the nonadiabatic coupling. For this goal, we add a nonzero $\gamma$ term in the Hamiltonian \eqref{H}, and we obtain
\be\label{HNH}
\H^\gamma(t) = \frac\hbar 2 \left[ \begin{array}{cc} \i\gamma\t  & \Omega (t) \\
\Omega(t) & 2\Delta\t -\i\gamma\t
\end{array} \right].
\ee
Two-level non-Hermitian Hamiltonians have been considered in several recent works (see, for instance, \cite{LZ-NH,uff1,uff2,uff3,uff4}) and used to model, for example, the dynamics of open two-level  systems or light transport in an optical directional coupler with gain and/or loss regions.  For example, the Hamiltonian \eqref{HNH} can be used to describe the physics of two coupled waveguides (cavities) with an asymmetric gain-loss profile and propagation constant detuning that varies with distance (time) \cite{uff2}.\\
In the basis $\ket{\varphi_\pm(t)}$ the Hamiltonian \eqref{HNH} has the form
\begin{align}\notag
&\mathbf{H}_a^\gamma\t = \\ 
&\hbar \left[ \begin{array}{cc}
 \lambda_{-}\t + \half\i\gamma\t\cos 2\theta\t  &  \frac12\i\gamma\t\sin2\theta\t -\i\dot{\theta}\t \\
  \frac12\i\gamma\t\sin2\theta\t + \i\dot{\theta}\t  & \lambda_{+}\t - \half\i\gamma\t\cos 2\theta\t
  \end{array} \right] .
\def\t{(t)}
\end{align}
where $\lambda_{\pm}(t)$ are again defined by Eq.~\eqref{eigenenergies}. We assume that initially the system is in state $\ket{1}$, which in that moment coincides with $\ket{\varphi_+(t)}$. Next, if we choose $\gamma\t=2\dot{\theta}\t/\sin2\theta\t$, we can nullify $H_{12}\t$, which means that state $\ket{\varphi_-(t)}$, which is not populated initially, never receives any population during the evolution. Since state $\ket{\varphi_+(t)}$ initially coincides with state $\ket{1}$, and finally with state $\ket{2}$, this allows the transfer to be realized at any arbitrary speed.

%%%%%%%%%%%%%%%%%%%%%%%%%%%%%%%%%%%%%%%%%%%%%%%%%%%%%%%%%%%%%%%%%%%%%%%%%%%%%%%%%%%%%%%%%%%%%%%%%%%%%%%%%%%%%%%%%%%%%%%
\begin{figure}[tb]
\centering \includegraphics[width=1.6\columnwidth]{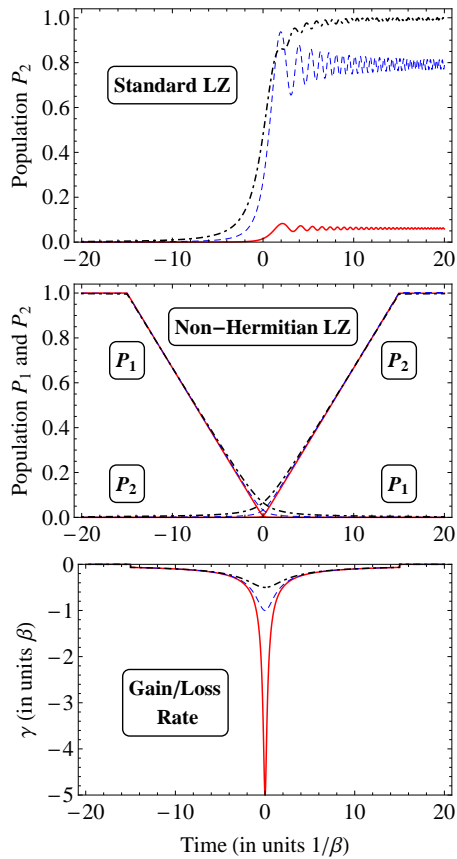}
\caption{(Color online) Time evolution of the populations for the standard LZ model (top frame) and with the addition of the non-Hermitian term (middle frame). Loss/gain rate as a function of time (bottom frame). The values of the interaction parameters are $\Omega_0/\beta=0.2$ (solid line), $\Omega_0/\beta=1$ (dashed line) and $\Omega_0/\beta=2$ (dot-dashed line).
}\label{LZfig}
\end{figure}
%%%%%%%%%%%%%%%%%%%%%%%%%%%%%%%%%%%%%%%%%%%%%%%%%%%%%%%%%%%%%%%%%%%%%%%%%%%%%%%%%%%%%%%%%%%%%%%%%%%%%%%%%%%%%%%%%%%%%%%

We shall now consider two special cases, which will reveal how to apply the described technique in a concrete situation. The first example is the LZ model,
\be
\Omega\t =\Omega_0 = \const,\quad \Delta\t = \beta^2 t ,
\ee
where $\beta^2$ is the slope of the crossing and, without loss of generality, we consider $\Omega_0 > 0$. It is convinient to introduce the dimensionless parameters $T=\beta t$ and $\omega=\Omega_0/\beta$. Then, it is easy to show that the nonadiabatic coupling has the form of a Lorentzian,
\be
\dot{\theta}\t = -\frac{\Omega_0\beta^2}{2(\Omega_0^2 + \beta^4 t^2)} = \frac{-\Omega_0}{2(\omega^2+T^2)}
\ee
and hence, in order to nullify it, we choose
\be\label{gamma}
\gamma\t=\frac{2\dot{\theta}\t}{\sin2\theta\t}=\frac{-1}{\sqrt{\Omega_0^2 / \beta^4  + t^2}} =\frac{-\beta}{\sqrt{\omega^2+T^2}}.
\ee
In such a way, the Hamiltonian in the basis $\ket{\varphi_\pm(t)}$ becomes
\begin{align}\notag
&\mathbf{H}_a^\gamma\t = \\ 
&\hbar \left[ \begin{array}{cc}
 \lambda_{-}\t + \half\i\gamma\t\cos 2\theta\t  &  0  \\
 2\i\dot{\theta}\t  & \lambda_{+}\t - \half\i\gamma\t\cos 2\theta\t
  \end{array} \right] ,
\def\t{(t)}
\end{align}
where
\be
\lambda_{\pm} =\frac{\beta}{2}\left(T\pm\sqrt{\omega^2+T^2}\right)
\ee
and
\be\label{cos}
\cos 2\theta = \frac{\omega T}{\sqrt{\omega^2 T^2+\omega^4}} .
\ee
If our system is prepared initially in state $\c(t_i)=[\sin\theta(t_i),\cos\theta(t_i)]\approx [1,0]$, then we will have $a_-(t_i)=0$ and $a_+(t_i)=1$. By using the Schr\"{o}dinger equation \eqref{Schr-adiabatic} we obtain for the evolution of the amplitudes
\bse
\begin{align}
&a_-(t_f)= 0 \\ \label{ap}
&a_+(t_f)=\exp \left(-\i\int_{t_i}^{t_f}\lambda_{+}\t - \half\i\gamma\t\cos 2\theta\t d t \right)
\end{align}
\ese
for any value of $t_f>t_i$.
It can be seen from Eq. \eqref{gamma} that $\gamma\t$ is an even function of time and from Eq. \eqref{cos} that $\cos2\theta\t$ is an odd function of time. Hence, if we assume that the $t_f=-t_i$, the norm of the state vector at $t_f$ is equal to unity, because the real part of the integral in Eq. \eqref{ap} is zero. This property holds whenever $\Omega\t$ is an even function of time and $\Delta\t$ is an odd function of time. In Fig.~\ref{LZfig} we compare the evolution of the populations $P_1$ and $P_2$ of the two bare states for the standard LZ model and for the one with the additional non-Hermitian term $\i\gamma$. It can be seen that in the case of the non-Hermitian LZ model the population transfer is always perfect, regardless of the value $\omega$ of the interaction strength. However, the smaller the value of $\omega$, the larger loss/gain rate has to be included. It should be noted here, that since the Hamiltonian is non-Hermitian, the norm of the state vector, given by $\sqrt{P_1+P_2}$, does not need to be conserved during the evolution. This property can be seen in Fig.~\ref{LZfig}(middle frame), where the norm is not conserved during the interaction. However, because of the symmetry of $\Omega$ and $\Delta$, the initial and final norm of the state vector is unity. Another important point that should be emphasized is that we only consider evolution in a finite time. Since the integral of $\gamma\t$ is divergent, we have to cut it in time in order to prevent the population $P_1$ to increase to values larger than unity. In the case of Fig.~\ref{LZfig}, the time interval is $T\in[-15,15]$. Before $t_i$ and after $t_f$, $\gamma$ is assumed equal to zero. We want to note here, that unlike the standard LZ model, where it is well known that the two bare energies cross in time, in the non-Hermitan LZ model, because of the extra imaginary term, the two curves do not cross in the complex plane. In Fig.~\ref{energies} we show a schematic plot of the bare energies $\varepsilon$ for the standard and non-Hermitian LZ models and this feature is well illustrated.

%%%%%%%%%%%%%%%%%%%%%%%%%%%%%%%%%%%%%%%%%%%%%%%%%%%%%%%%%%%%%%%%%%%%%%%%%%%%%%%%%%%%%%%%%%%%%%%%%%%%%%%%%%%%%%%%%%%%%%%
\begin{figure}[tb]
\centering \includegraphics[width=0.9\columnwidth]{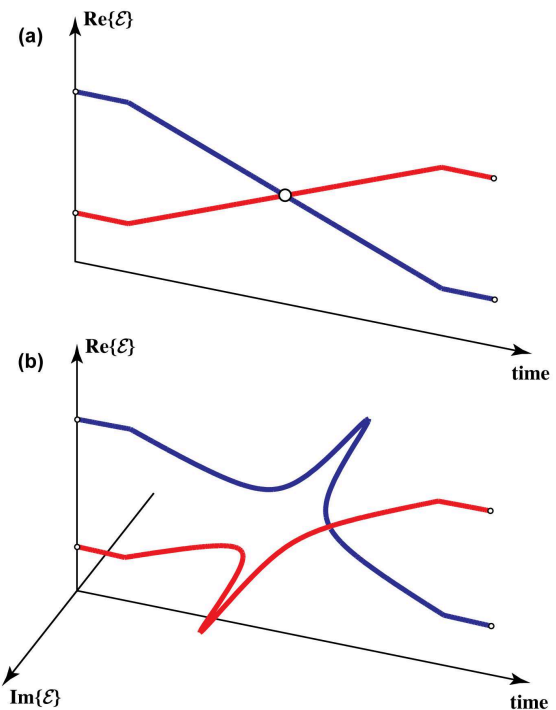}
\caption{(Color online) Bare energies for the LZ model. (a) Standard LZ model, the energies cross in time. (b) Non-Hermitian LZ model, because of the imaginary term $\i\gamma\t$, the energies do not cross.}\label{energies}
\end{figure}
%%%%%%%%%%%%%%%%%%%%%%%%%%%%%%%%%%%%%%%%%%%%%%%%%%%%%%%%%%%%%%%%%%%%%%%%%%%%%%%%%%%%%%%%%%%%%%%%%%%%%%%%%%%%%%%%%%%%%%%

It is worth saying that a physical implementation of the non-Hermitian shortcut to adiabaticity above designed for the LZ model can be accomplished in waveguide optics. As reported in Ref.~\cite{Dreisov_PRA_09}, LZ tunneling with linear crossing of energy levels can be mimicked in a directional coupler with a cubically bent profile for the waveguide axis. The required gain/loss imbalance between the two waveguides $2i\gamma(t)$, with $\gamma(t)$ provided by Eq.~\eqref{gamma}, can be implemented by cascading a purely dissipative coupler with non-uniform propagation loss, and an active coupler with uniform gain, precisely as suggested in a recent paper on PT-symmetric unidirectional reflectionless metamaterials~\cite{Feng_NM_13}. Finally, the non-uniform loss profile $\gamma(t)$ can be obtained by evanescent coupling of the waveguide mode with a metallic thin film cover of suitable geometry deposited on top of the passive waveguides, a technique that has been already exploited to produce a sinusoidally-shaped loss profile along the axis of a silicon on silica channel waveguide~\cite{Feng_NM_13}.

\begin{figure}[tb]
\centering \includegraphics[width=1.6\columnwidth]{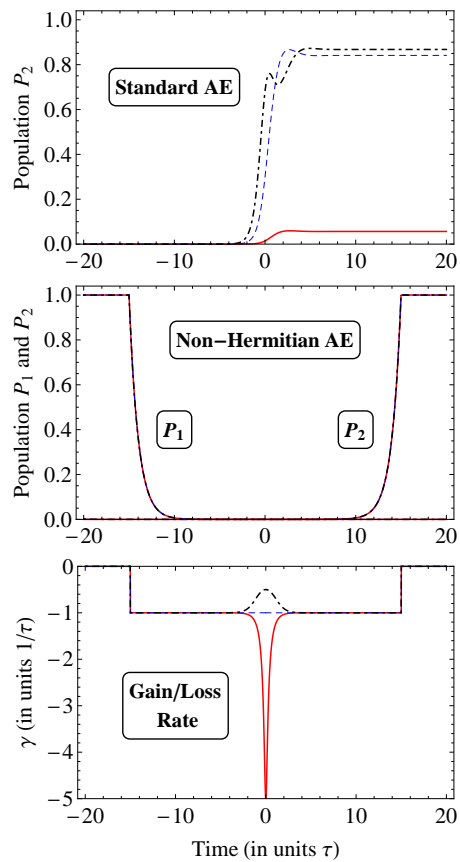}
\caption{(Color online)Same as Fig.~\ref{LZfig}, but for the AE model. The values of the interaction parameters are $B\tau=1$ and $\Omega_0\tau=0.2$ (solid line), $\Omega_0\tau=1$ (dashed line) and $\Omega_0\tau=2$ (dot-dashed line). Unlike for the non-Hermitian LZ model, here the three curves for $P_1$ and $P_2$ in the non-Hermitian AE model are undistinguishable.
}\label{AEfig}
\end{figure}

As a second example, we consider the AE model
\be
\Omega\t =\Omega_0 \sech (t/\tau) ,\quad \Delta\t = D \tanh (t/\tau) ,
\ee
where $\tau$ is the characteristic duration of the interacion and $D$ is a real parameter, corresponding to the chirp rate. We proceed the same way as with the LZ model and calculate
\be
\dot{\theta}\t =\frac{1}{\tau}\frac{\delta\alpha\cosh(t/\tau)}{\delta^2 -2\alpha^2-\delta^2\cosh(2t/\tau)},
\ee
where $\alpha =\Omega_0 \tau$ and $\delta=D\tau$. For the gain/loss function we obtain
\be
\gamma\t =\frac{1}{\tau}\frac{-2\delta(\e^{t/\tau}+\e^{3t/\tau})}{(\e^{2t/\tau}-1)^2\sqrt{\csch^2 (t/\tau)(\delta^2+\alpha^2\csch^2 (t/\tau))}},
\ee
where again, like for the LZ model, we assume that this function is taken within some finite symmetric time interval. In Fig.~\ref{AEfig} we show the population evolution for the AE model with and without the addition of the term $\i\gamma$. Unlike the LZ model, here the function $\gamma$ does not vanish at $\pm\infty$, but tends to a constant value. 

%%%%%%%%%%%%%%%%%%%%%%%%%%%%%%%%%%%%%%%%%%%%%%%%%%%%%%%%%%%%%%%%%%%%%%%%%%%%%%%%%%%%%%%%%%%%%%%%%%%%%%%%%%%%%%%%%%%%%%%
%%%%%%%%%%%%%%%%%%%%%%%%%%%%%%%%%%%%%%%%%%%%%%%%%%%%%%%%%%%%%%%%%%%%%%%%%%%%%%%%%%%%%%%%%%%%%%%%%%%%%%%%%%%%%%%%%%%%%%%

\section{Discussion and conclusion}

In this work we have proposed a method of non-Hermitian shortcut to adiabaticity, that enables to achieve an arbitrarily fast population transfer in a two-state quantum system. This is performed by introducing a non-Hermitian term in the Hamiltonian, which has the purpose to cancel the nonadiabatic coupling. The improvement of the population transfer is explicitly demonstrated for the special cases of the Landau-Zener and Allen-Eberly models.  A few major differences between our method and the  shortcut technique of Hermitian systems should be highlighted.  In the standard Hermitian shortcuts to adiabaticity \cite{shortcuts}  additional fields are used, which couple the bare states in such way that the resultant nonadiabatic coupling is zero. These techniques allow to speed up the adiabatic evolution, but at the cost of increasing the coupling. Conversely, in the non-Hermitian case proposed in our work  the population transfer can be made arbitrarily fast, even for an arbitrarily small coupling between the two states, by introducing a complex-valued detuning (energies) of the uncoupled system. This non-Hermitian term corresponds to gain or loss (depending on the sign) of population in the two bare states. Such terms can be physically realized, for instance, in two coupled optical waveguides with longitudinally-varying gain and loss regions \cite{uff2}. 
A second feature is that the shortcut to adiabaticity in the non-Hermitian model is sensitive to the initial condition of the system. In order to have the norm of the state vector preserved, we need to start exactly from $\c(t_i)=[\sin\theta(t_i),\cos\theta(t_i)]$, which in our case is approximately equal to $[1,0]$. If we deviate from this condition, the technique will still produce complete transfer of population, but without preserving the norm and some extra gain or loss may be introduced. Finally, a third and very distinctive difference is  that the non-Hermitian shortcut to adiabaticity is not symmetric against flipping the initial condition. As the standard Hermitian shortcuts produce complete population transfer both for the initial conditions $\c(t_i)=[1,0]$ and $\c(t_i)=[0,1]$,  our technique demands a change in the sign of $\gamma$ (i.e. the interchange of gain and loss terms) for the two different situations.  Our analysis suggests that adiabatic passage techniques well-developed for Hermitian systems can be extended to non-Hermitian ones, with extra degrees of freedoms and novel dynamical features.  It is envisaged that our study, which has been focused to the simplest cases of adiabatic passage methods in a two-level system, could stimulate further studies of coherent population transfer techniques of multi-level systems (such as STIRAP \cite{NVV-adiabatic}) for non-Hermitian systems. 
\par
%%%%%%%%%%%%%%%%%%%%%%%%%%%%%%%%%%%%%%%%%%%%%%%%%%%%%%%%%%%%%%%%%%%%%%%%%%%%%%%%%%%%%%%%%%%%%%%%%%%%%%%%%

\section{Acknowledgement}
This work was supported by the Fondazione Cariplo (Grant No. 2011-0338).

%%%%%%%%%%%%%%%%%%%%%%%%%%%%%%%%%%%%%%%%%%%%%%%%%%%%%%%%%%%%%%%%%%%%%%%%%%%%%%%%%%%%%%%%%%%%%%%%%%


\begin{thebibliography}{99}

\bibitem{NVV-adiabatic} N.V. Vitanov et al., Annu. Rev. Phys. Chem. \textbf{52}, 763 (2001).

\bibitem{CAP} B. T. Torosov, S. Gu\'{e}rin, and N. V. Vitanov, Phys. Rev. Lett. \textbf{106}, 233001 (2011).

\bibitem{parallel}
S. Gu\'{e}rin, S. Thomas, and H.R. Jauslin, Phys. Rev. A {\bf 65}, 023409 (2002);
%S. Gu\'{e}rin \etal, Phys. Rev. A \textbf{65}, 023409 (2002);
%G. Dridi, S. Gu\'{e}rin, V. Hakobyan, H. R. Jauslin, and H. Eleuch, \ibid \textbf{80}, 043408 (2009).
G. Dridi \etal, \ibid \textbf{80}, 043408 (2009).


\bibitem{shortcuts}
%R. G. Unanyan, L. P. Yatsenko, K. Bergmann, and B. W. Shore, Opt. Commun. \textbf{139}, 48 (1997);
R. G. Unanyan \etal, Opt. Commun. \textbf{139}, 48 (1997);
 M. V. Berry, J. Phys. A \textbf{42}, 365303 (2009);
 Xi Chen, I. Lizuain, A. Ruschhaupt, D. Gu{\'e}ry-Odelin, and J.G. Muga, Phys. Rev. Lett. \textbf{105}, 123003 (2010).

%\bibitem{Chen10} Xi Chen, I. Lizuain, A. Ruschhaupt, D. Gu{\'e}ry-Odelin, and J.G. Muga, Phys. Rev. Lett. \textbf{105}, 123003 (2010).

% Shortcut to Adiabatic Passage in Two- and Three-Level Atoms

%\bibitem{Chen12} Xi Chen and J.G. Muga, Phys. Rev. A \textbf{86}, 033405 (2012).

\bibitem{shortcuts-NH}
S. Ib\'{a}\~{n}ez, S. Martinez-Garaot, Xi Chen, E. Torrontegui, and J. G. Muga,
Phys. Rev. A \textbf{84}, 023415 (2011)

\bibitem{PT} C. M. Bender and S. Boettcher, Phys. Rev. Lett. \textbf{80}, 5243 (1998);
C. M. Bender, Rep. Prog. Phys. {\bf 70}, 957 (2007).

\bibitem{Faster} C.M. Bender, D.C. Brody, H.F. Jones, and B.K. Meister, Phys. Rev. Lett. \textbf{98}, 040403 (2007);
R. Uzdin, U. G\"{u}nther, S. Rahav, and N. Moiseyev, J. Phys. A: Math. Theor. \textbf{45}, 415304 (2012).

\bibitem{LZ-NH} E.M. Graefe, H.J. Korsch, Czech. J. Phys. \textbf{56}, 1007 (2006); 
S. A. Reyes, F. A. Olivares and L. Morales-Molina, J. Phys. A: Math. Theor. \textbf{45}, 444027 (2012); 
R. Uzdin and N. Moiseyev, J. Phys. A: Math. Theor. \textbf{45}, 444033 (2012).

\bibitem{uff1}
N. Moiseyev, Phys. Rev. A {\bf 83}, 052125 (2011).

\bibitem{uff2}
R. El-Ganainy, K.G. Makris, and D.N. Christodoulides, Phys. Rev. A {\bf 86}, 033813 (2012).

\bibitem{LZ} L. D. Landau, Physik Z. Sowjetunion \textbf{2}, 46 (1932); C. Zener, Proc. R. Soc. Lond. Ser. A \textbf{137}, 696 (1932);
E. C. G. St\"{u}ckelberg, Helv. Phys. Acta \textbf{5}, 369 (1932); E. Majorana, Nuovo Cimento \textbf{9}, 43 (1932).

\bibitem{AE} L. Allen and J.H. Eberly \emph{Optical Resonance and Two-Level Atoms} (New York, Dover, 1987).

\bibitem{GSV} G. S. Vasilev and N. V. Vitanov, J. Chem. Phys. \textbf{123}, 174106 (2005)

\bibitem{uff3}
E. M. Graefe, H. J. Korsch, and A. E. Niederle, 
Phys. Rev. Lett. {\bf 101}, 150408 (2008); E.-M. Graefe, H.J. Korsch, and A.E. Niederle, Phys. Rev. A {\bf 82}, 013629 (2010).

\bibitem{uff4}
K. Xiao, W. Hai, and J. Liu, Phys.Rev. A {\bf 85}, 013410 (2012).

\bibitem{Dreisov_PRA_09} F. Dreisow, A. Szameit, M. Heinrich, S. Nolte, A. T\"{u}nnermann, M. Ornigotti and S. Longhi, Phys. Rev. A \textbf{79}, 055802 (2009).

\bibitem{Feng_NM_13} L. Feng, Y.-L. Xu, W. S. Fegadolli, M.-H. Lu, J. E. B. Oliveira, V. R. Almeida, Y.-F. Chen, and A. Scherer, Nature Mater. \textbf{12}, 108 (2013).


\end{thebibliography}
\end{document}